\documentclass[aps,prb,reprint,twocolumn,a4paper,showpacs,superscriptaddress,groupedaddress]{revtex4-1} 
\usepackage{amsmath}
\usepackage{amssymb}
\usepackage{bm}
\usepackage[dvips]{graphicx} 

\def\bra#1{\mathinner{\langle{#1}|}}
\def\ket#1{\mathinner{|{#1}\rangle}}
\def\braket#1{\mathinner{\langle{#1}\rangle}}

\begin{document}
  \title{Time-dependent many-variable variational Monte Carlo method \\for nonequilibrium strongly correlated electron systems}
  \author{Kota Ido, Takahiro Ohgoe and Masatoshi Imada}
  \affiliation{Department of Applied Physics, University of Tokyo, 7-3-1 Hongo, Bunkyo-ku, Tokyo 113-8656, Japan}

\begin{abstract}
We develop a time-dependent variational Monte Carlo (t-VMC) method for quantum dynamics of strongly correlated electrons. 
The t-VMC method has been recently applied to bosonic systems and quantum spin systems. 
Here, we propose a time-dependent trial wave function with many variational parameters, which is suitable for nonequilibrium strongly correlated electron systems.
As the trial state, we adopt the generalized pair-product wave function with correlation factors and quantum-number projections.
This trial wave function has been proven to accurately describe ground states of strongly correlated electron systems.
To show the accuracy and efficiency of our trial wave function in nonequilibrium states as well, we present our benchmark results for relaxation dynamics during and after interaction quench protocols of fermionic Hubbard models. 
We find that our trial wave function well reproduces the exact results for the time evolution of physical quantities such as energy, momentum distribution, spin structure factor, and superconducting correlations. 
These results show that the t-VMC with our trial wave function offers an efficient and accurate way to study challenging problems of nonequilibrium dynamics in strongly correlated electron systems.
\end{abstract}

\maketitle
\section{Introduction}
Quantum systems with strong many-body correlations in equilibrium show intriguing properties such as the metal-insulator transition\cite{RevModPhys.70.1039} and high-temperature superconductivity\cite{bednorz1986possible,kamihara2008iron}. Recently, because of potential routes to realizing intriguing phenomena that are not attainable in the equilibrium, strongly correlated electron systems driven out of equilibrium have attracted much attention. In fact, owing to the development of experimental techniques, we have been able to control or realize unprecedented phases and their phase transitions by applying strong and short pulse of external fields such as intensive laser pumping\cite{koshihara1990photoinduced,PhysRevLett.111.187801,Rini2007,Okamoto2010, Okamoto2011,Ichikawa2011, Stojchevska2014,Fausti2011, Hu2014, Kaiser2014}.

For satisfactory theoretical understanding of quantum dynamics of many-body systems, we have to solve many-body time-dependent Schr\"odinger equation $i\frac{d}{dt} \ket{\psi(t)} = \mathcal{H}(t)\ket{\psi(t)}$, where $\ket{\psi(t)}$ and $\mathcal{H}(t)$ represent the wave function and the Hamiltonian at time {\it t}, respectively. The formal solution of the time-dependent Schr\"odinger equation is given by $\ket{\psi(t)}=\mathcal{T}\exp(-i \int_0^t \mathcal{H}(s)ds ) \ket{\psi(0)} $. Here, $\mathcal{T}$ represents the time ordering. 
 However, such an approach is tractable only for small many-body systems, because the Hilbert space grows exponentially as the system size increases. Furthermore, reduction to an effective single-particle problems such as the time-dependent Hartree-Fock method \cite{dirac1930time} does not give us accurate results. To treat larger systems accurately, there exist several numerical methods such as time-dependent density matrix renormalization group method(DMRG) \cite{PhysRevLett.93.040502, PhysRevLett.93.076401, 1742-5468-2004-04-P04005}, and nonequilibrium dynamical mean field theory(DMFT) \cite{RevModPhys.86.779} and quantum Monte Carlo(QMC) method \cite{Goth2012}. 
 However, DMRG and DMFT have difficulties in treating large systems in two or three spatial dimensions when one wishes to treat spatial correlations and fluctuations accurately.
 In order to include non-local correlations, the dynamical cluster approximation(DCA) has been proposed as an extension of the DMFT\cite{Tsuji2014}. 
 However, it requires high computational costs for a large cluster size.
 Although the QMC method can treat finite systems exactly, its applications are very limited due to the notorious negative-sign problem.
Recently, to overcome the above difficulties, Carleo {\it et al.} developed the time-dependent variational Monte Carlo (t-VMC) method 
 and optimized Jastrow factors in bosonic systems\cite{carleo2012localization, PhysRevA.89.031602}.
This method can be formulated based on the time dependent variational principle (TDVP)\cite{dirac1930time, mclachlan1964variational,PhysRevB.88.075133}. 
The t-VMC method has been applied  not only to bosonic systems\cite{carleo2012localization, PhysRevA.89.031602} but also to spin models\cite{PhysRevA.92.041603}.
However, to the best of our knowledge, its application to correlated electron systems has not been successful yet. 
This limitation may be ascribed to the difficulty of constructing an accurate trial wave function for such systems. 
Therefore, the proposal of accurate trial wave functions suitable for nonequilibrium electron systems in the t-VMC method is desirable.

The purpose of this study is to propose an accurate and efficient trial wave function for strongly correlated electron systems out of equilibrium in the t-VMC framework. 
To achieve this purpose, we focus on highly accurate trial wave functions for ground states of correlated electron systems.
In such systems, many studies have attempted to construct an accurate and efficient trial wave function in the variational Monte Carlo (VMC) framework\cite{sorella2001generalized,PhysRevLett.94.026406,casula2004correlated,PhysRevB.78.041101,PhysRevB.85.045103,tahara2008variational}.
In Ref.[\onlinecite{tahara2008variational}], Tahara and one of the authors have reduced the biases by using the quantum-number projections and introducing many variational parameters to one-body part. 
They have adopted a generalized pair-product wave function as a one-body part because it can flexibly describe different competing phases such as correlated metals, antiferromagnetic states and superconducting states.
This improved trial wave function has proven to be highly accurate for ground states of strongly correlated electron systems\cite{tahara2008variational,kaneko2013improved,PhysRevB.90.115137}.
In this paper, we show that this trial wave function is an accurate and efficient one even for nonequilibrium strongly correlated electron systems in the t-VMC framework.

The organization of this paper is as follows. 
In Sec. I\hspace{-.1em}I, we introduce the TDVP which enables us to obtain an optimal time-dependent trial wave function and formulate the t-VMC method by using the TDVP.
Section I\hspace{-.1em}I\hspace{-.1em}I describes a trial wave function with a large number of variational parameters for nonequilibrium strongly correlated electron systems. 
In Sec. I\hspace{-.1em}V, we show the accuracy and efficiency of our trial wave function by presenting several benchmark results. 
Finally, we summarize our work in Sec. V.
\section{time-dependent variational principle}
The time-dependent variational principle (TDVP) proposed by McLachlan is a variational principle for time-dependent wave functions \cite{mclachlan1964variational}. In this principle, we consider a distance between $i\frac{d}{dt} \ket{\psi_{\bm{\alpha}}}$ and $\mathcal{H}\ket{\psi_{\bm{\alpha}}}$ where $\bm{\alpha}=\{ \alpha_k | k=1, \cdots, N_p\}$ represent time-dependent variational parameters. By definition, the distance satisfies the inequality
\begin{eqnarray}
\underset{\bm{\alpha}}{\rm min} \left\| i\frac{d}{dt} \ket{\psi_{\bm{\alpha}}} - \mathcal{H}\ket{\psi_{\bm{\alpha}}}\right\| \geq 0, \label{tdvp0}
\end{eqnarray}
where the equality holds if $\ket{\psi_{\bm{\alpha}}}$ is the solution of the time-dependent Schr\"odinger equation. 
Here, the norm $\| \ket{\Psi} \|$ is defined as the square root of an inner product of a wave function $\ket{\Psi}$, i.e. $\| \ket{\Psi} \|=\sqrt{\braket{\Psi | \Psi}}$.
If we could optimize the variational parameters at each time step such that the equality holds, we obtain the exact solution of $\ket{\psi_{\bm{\alpha}}}$. 
If a trial wave function well approximates the exact solution of the time-dependent Schr\"odinger equation, the value of the lower bound should be small. Based on this idea, we optimize variational parameters at each time-step such that the distance is minimized. Originally, the TDVP was applied in the field of quantum chemistry \cite{heller1976time, beck2000multiconfiguration}. Recently, a similar principle has been applied to the matrix product state for quantum spin models \cite{haegeman2011time, PhysRevB.88.075133}, the bosonic Jastrow-type wave function for the Bose-Hubbard model \cite{carleo2012localization, PhysRevA.89.031602} and the Gutzwiller approximation for strongly correlated electron systems \cite{PhysRevLett.105.076401, PhysRevB.83.165105, PhysRevB.91.115102}.

Although exact time evolution is unitary, and thus, the norm $\braket{\psi_{\bm{\alpha}} | \psi_{\bm{\alpha}}}$ is conserved, it is not necessary conserved in TDVP [Eq. (\ref{tdvp0})]. To remove the restriction on the norm, we use a TDVP for norm-independent dynamics\cite{PhysRevB.88.075133},
\begin{eqnarray}
\underset{\bm{\alpha}}{\rm min} \left\| \left(1-\frac{\ket{\psi_{\bm{\alpha}}} \bra{\psi_{\bm{\alpha}}}}{\braket{\psi_{\bm{\alpha}}| \psi_{\bm{\alpha}}}} \right) \left[ i\frac{d}{dt} \ket{\psi_{\bm{\alpha}}} - \mathcal{H}\ket{\psi_{\bm{\alpha}}} \right]\right\|  \geq 0. \label{tdvp}
\end{eqnarray}
The details of the TDVP for norm-independent dynamics is described in Appendix \ref{detail}.
Based on this TDVP, we can derive the differential equation of the time-dependent variational parameters. 
Namely, by solving the minimization problem on the distance (\ref{tdvp}), we obtain the time evolution of the variational parameters\cite{haegeman2011time,PhysRevB.88.075133, carleo2012localization,PhysRevA.89.031602}:
\begin{eqnarray}
\dot{\alpha_k} =\frac{d \alpha_k}{dt}= -i \sum_l^{N_p} (S^{-1})_{kl}g_l, \label{tdvp_eq}
\end{eqnarray}
where a matrix $S$ and a vector $g$ are described as
\begin{eqnarray}
&S_{kl} = \braket{ \mathcal{O}^\dagger_k \mathcal{O}_l} - \braket{ \mathcal{O}^\dagger_k}\braket{ \mathcal{O}_l}, \\
&g_k = \braket{ \mathcal{O}^{\dagger}_k \mathcal{H}} - \braket{\mathcal{O}_k^\dagger} \braket{\mathcal{H}}, 
\end{eqnarray}
respectively. In the t-VMC method, we estimate an expectation value $\braket{A} = \frac{\braket{\psi_{\bm{\alpha}} | A | \psi_{\bm{\alpha}}}}{\braket{\psi_{\bm{\alpha}} | \psi_{\bm{\alpha}}}}$ by the Markov-chain Monte Carlo method. 
The derivative operators $\mathcal{O}_k$ and $\mathcal{O}_k^\dagger$ are defined by using real space configurations of electrons $\{x\}$ as
\begin{eqnarray}
\left. \begin{array}{l}
\mathcal{O}_k={\displaystyle \sum_{x}} \ket{x} O_k(x) \bra{x},  \\
\mathcal{O}^\dagger_k={\displaystyle \sum_{x}} \ket{x} O^{*}_k(x) \bra{x}, 
\end{array}\right.
\end{eqnarray}
respectively. Here,
\begin{eqnarray}
\left. \begin{array}{l}
O_k(x) = {\displaystyle \frac{1}{\braket{x | \psi_{\bm{\alpha}}}} \frac{\partial}{\partial \alpha_k} \braket{x| \psi_{\bm{\alpha}}}, } \\
O^*_k(x) = {\displaystyle \frac{1}{\braket{\psi_{\bm{\alpha}} | x}} \frac{\partial}{\partial \alpha_k^*} \braket{\psi_{\bm{\alpha}} | x} }.
\end{array}\right.
\end{eqnarray}
The differential equation [Eq. (\ref{tdvp_eq})] is called {\it TDVP equation} \cite{haegeman2011time, PhysRevB.88.075133}. This TDVP equation can also be derived by minimizing the time-dependent action \cite{PhysRevB.88.075133,kramer1981geometry}. If we use the time-dependent variational principle for imaginary time evolution $t=-i\tau$ and solve TDVP equation by using Euler method, we obtain the stochastic reconfiguration scheme proposed by Sollera\cite{sorella2001generalized}.
The TDVP equation has a symplectic property \cite{PhysRevB.88.075133,kramer1981geometry,hairer2006geometric}. This property leads to the energy conservation if the Hamiltonian is time-independent and we could calculate the derivative of parameters $\dot{\alpha_k}$ exactly.

In this study, in order to solve the TDVP equation, we use the fourth-order Runge-Kutta method which provides us with a stable and efficient way to perform the time integration. 
Note that the Runge-Kutta method is not a symplectic integral method. 
Furthermore, there are stochastic errors in the Monte Carlo calculation of quantities such as $\braket{\mathcal{O}^\dagger_k \mathcal{H} }$ and $\braket{\mathcal{O}^\dagger_k \mathcal{O}_l }$. 
These cause the breaking of the symplectic property of the TDVP equation. Nevertheless, we observed that the energy is conserved with high accuracy as we show in Sec. I\hspace{-.1em}V.

\section{time-dependent variational Wave Function}
\label{twf}
In the t-VMC method, the choice of trial wave functions is important.
As a trial wave function, we adopt the form of
\begin{eqnarray}
\ket{\psi(t)} = \mathcal{L} \mathcal{P}(t) \ket{\phi(t)},
\end{eqnarray}
which has been used for equilibrium systems in Refs.[\onlinecite{tahara2008variational,kaneko2013improved,PhysRevB.90.115137}]. 
Here, $\mathcal{L}$ represents quantum-number projections which recovers the symmetries the wave function should have throughout the time evolution, and $\mathcal{P}(t)$ represents correlation factors. For the one-body part $\ket{\phi(t)}$, we employ the pair-product wave function. In addition, we include backflow correlations in the pair-product wave function for lattice model \cite{PhysRevB.78.041101, PhysRevB.83.195138}. 
The time-dependent variational parameters are included in the correlation factors as well as in the one-body part with the backflow correlations.
Note that these variational parameters should be treated as {\it complex} numbers because the variational parameters evolve as complex numbers in the present method. 
In this section, we describe each component in detail.

\subsection{One-body Part}
In the conventional VMC, the Slater determinant with small variational parameters is used as the one-body part.
In order to improve the conventional one-body part, we assume the form of the pair-product wave function with many variational parameters \cite{tahara2008variational}:
\begin{eqnarray}
\ket{\phi} = \left( \sum_{i,j}^{N_s} f_{ij} c^\dagger_{i\uparrow} c^\dagger_{j\downarrow} \right)^{N/2}\ket{0},\label{onebody} 
\end{eqnarray}
where $N$ is the number of electrons, $N_s$ is the system size, $c^\dagger_{i\sigma}$ ($c_{i\sigma}$) is a creation (annihilation) operator of an electron with spin $\sigma$ on the site $i$, and the pairing amplitude $f_{ij}$ is treated as variational parameters.
This pair-product wave function is a general form of a Hartree-Fock-Bogoliubov-type wave function which allows antiferromagnetic and superconducting orders\cite{tahara2008variational,giamarchi1991phase, himeda2000spontaneous}.
Thus, it takes an advantage of flexibly describing the paramagnetic metals, antiferromagnetic ordered states and superconducting states with any type of frequency independent gap on equal footing.
\subsection{Correlation Factors}
By operating the correlation factors on the pair-product wave function, we can include many-body correlation effects beyond the mean-field level. In this study, we use the Gutzwiller factor $\mathcal{P}_G$\cite{gutzwiller1964phys} and the Jastrow factor $\mathcal{P}_J$\cite{jastrow1955many}, i.e., $\mathcal{P} = \mathcal{P}_G \mathcal{P}_J$.

The Gutzwiller factor which was introduced by Gutzwiller \cite{gutzwiller1964phys} has the form of
\begin{eqnarray}
\mathcal{P}_G = \exp \left( -g\sum_i^{N_s} n_{i\uparrow} n_{i\downarrow} \right),
\end{eqnarray}
where $n_{i\sigma} = c^\dagger_{i\sigma}c_{i\sigma}$ and $g$ is the variational parameter. The Gutzwiller factor punishes the double occupation of electrons on the same site in real space configurations. In the limit $g \rightarrow \infty$, the Gutzwiller wave function $\mathcal{P}_G\ket{\phi}$ corresponds to a state which contains no double occupation.
The Gutzwiller factor is a simple way to improve a mean-field wave function such as a Slater determinant and the pair-product wave function. 
However, it was numerically proven that the Gutzwiller wave function cannot describe the nonmagnetic Mott transition in any finite dimensional systems\cite{yokoyama1987variational}. The main reason for this is that the Gutzwiller factor only includes the on-site correlation. Although some doubly-occupied (doublon) and empty (holon) sites exit in the Mott insulator where charge fluctuations are allowed, the doublon and holon have to be bound in realizing an insulating behavior. 

In order to describe the Mott transition, 
the Jastrow factor is introduced\cite{PhysRevLett.94.026406}: 
\begin{eqnarray}
\mathcal{P}_J = \exp \left( -\sum_{i,j}^{N_s} v_{ij} (n_{i}-1) (n_{j}-1) \right),
\end{eqnarray}
where $n_i=n_{i\uparrow}+n_{i\downarrow}$, $v_{ij}=v(\bm{r}_i-\bm{r}_j)=v(\bm{r}_j-\bm{r}_i)$ are the variational parameters, and $\bm{r}_i (\bm{r}_j)$ represents the position vector of the site $i (j)$. 
The Jastrow factor can be represented by using a doublon number operator and a holon number operator. 
The electron number operator $n_i$ is written as $n_i=1+D_i-H_i$, where $D_i=n_{i\uparrow}n_{i\downarrow}$ is the doublon number operator and $H_i=(1-n_{i\uparrow})(1-n_{i\downarrow})$ is the holon number operator. By using this relation, Jastrow factor becomes
\begin{eqnarray}
\mathcal{P}_J = \exp \left( -\sum_{i,j}^{N_s} v_{ij} (D_iD_j + H_i H_j - D_iH_j - H_iD_j) \right). \nonumber \\ \label{jas}
\end{eqnarray}
Equation (\ref{jas}) shows that the Jastrow factor includes off-site repulsive correlations between doublon-doublon (holon-holon) pairs and attractive correlations between doublon-holon pairs if $v_{ij} > 0$. This doublon-holon attractive correlations in the Jastrow factor play a crucial role for the Mott transition \cite{PhysRevLett.94.026406,JPSJ.80.084705}.

\subsection{Quantum-Number Projections}
In general ground states of finite quantum systems, the symmetries of the Hamiltonian must be preserved even if the symmetry-breaking occurs in the thermodynamic limit. Furthermore, such symmetries must be preserved even after time evolutions as long as external fields do not break the original symmetries. However, conventional trial wave functions often break symmetries of the Hamiltonian.

The quantum-number projection enables us to recover the symmetries of the trial wave function\cite{ring2004nuclear}.
Here, we introduce two quantum-number projections: the spin projection $\mathcal{L}^S$ and the momentum projection $\mathcal{L}^{K}$. 
The spin projection $\mathcal{L}^S$ is the projection onto the state with a total spin $S$ and $z$-component of spin $S^z=0$. This projection has a form of the integration over spin space:
\begin{eqnarray}
\mathcal{L}^S = \frac{2S+1}{8\pi^2} \int d\Omega P_S(\cos \beta) \mathcal{R}(\Omega),
\end{eqnarray}
where $\Omega=(\alpha, \beta, \gamma)$ is the Euler angle, $P_S(\cos \beta)$ is the $S$-th Legendre polynomial, and $\mathcal{R}(\Omega) = e^{i\alpha S^z}e^{i\beta S^y}e^{i\gamma S^z}$ is the rotational operator. 
When the one-body part $\ket{\phi}$ and the real space configurations $\{ x\}$ satisfy the condition of $S^z=0$, we can omit the integrations over $\gamma$ and $\alpha$ as follows:
\begin{eqnarray}
&&\mathcal{L^S}\ket{\phi} =  \frac{2S+1}{8\pi^2} \int d\Omega P_S(\cos \beta) e^{i\alpha S^z}e^{i\beta S^y}e^{i\gamma S^z} \ket{\phi} \nonumber \\
&&=  \sum_x \ket{x}  \frac{2S+1}{2} \int d\beta \sin \beta P_S(\cos \beta) \braket{x | e^{i\beta S^y} |\phi}. 
\end{eqnarray}
We can efficiently estimate the integration over $\beta$ by the Gauss-Legendre quadrature \cite{Press:1992:NRC:148286}.
The momentum projection $\mathcal{L}^{\bm{K}}$ is the projection onto the state with a total momentum $\bm{K}$. This projection has a form:
 \begin{eqnarray}
\mathcal{L}^{\bm{K}} = \frac{1}{N_s} \sum_{\bm{R}}e^{i\bm{K} \cdot \bm{R}} \mathcal{T}_{\bm{R}},
\end{eqnarray}
where ${\mathcal{T}}_{\bm{R}}$ is the translation operator with the translation vector $\bm{R}$.
\subsection{Backflow Correlations}
One efficient way to include correlation effects in a trial wave function is to include backflow correlations\cite{PhysRev.102.1189, PhysRevB.48.12037, PhysRevB.77.115112, PhysRevB.78.041101, PhysRevB.83.195138}. 
Recently, Tocchio {\it et al.} proposed a way of introducing backflow correlations into a Slater determinant for lattice models and found that the backflow correlations substantially improve ground-state energy of frustrated electronic systems in the region above intermediate strength of coupling \cite{PhysRevB.78.041101,PhysRevB.83.195138}. 
In a way similar to the approach by Tocchio {\it et al.}, the backflow correlations for lattice models can be implemented in the pair-product wave functions with the momentum projection as
\begin{eqnarray}
\mathcal{L}^{\bm{K}} \ket{\phi^b} &=& \sum_{x} \mathcal{L}^{\bm{K}} \ket{x} \braket{x | \phi^b} \nonumber \\ 
&=& \sum_{x}\left( \frac{N}{2} \right)! \sum_{\bm{R}} e^{i\bm{K} \cdot \bm{R}} {\rm Pf} \left[ X^b(x_{-\bm{R}}) \right]  \ket{x}, \qquad
\end{eqnarray}
where $\ket{\phi^b}$ and ${\rm Pf} X^b(x_{-\bm{R}})$ represent the pair-product wave function with backflow correlations and the Pfaffian of the skew-symmetric matrix $X^b(x_{-\bm{R}})$, respectively. $X^b(x_{\bm{R}})$ is defined as
\begin{eqnarray}
 X_{n m}^b(x_{\bm{R}}) = f_{T_{\bm{R}}(i_n) T_{\bm{R}}(i_m)}^b(x_{\bm{R}}) -  f_{T_{\bm{R}}(i_m) T_{\bm{R}}(i_n)}^b(x_{\bm{R}}).\qquad
\end{eqnarray}
Here, $x_{\bm{R}}$ represents the real space configuration which is created by shifting a configuration $x$ by a translation vector $\bm{R}$. 
The site of the $n$-th ($m$-th) electron is represented by $i_n$ ($i_m$). $T_{\bm{R}}(i_n)$ is the site characterized by the position vector $\bm{r}_{i_n}+\bm{R}$. 
For simplicity, we do not consider the momentum projection in the following equations. 
The pairing amplitude with backflow correlations is defined by
\begin{eqnarray}
&f_{i_n i_m}^b(x) = {\displaystyle \sum_{\mu,\nu=0}^{3} \sum_{{\tau},{\tau}'}} \eta^{\mu\nu}_{{\tau} {\tau}'} \Theta_{{i_n},{i_n}+\tau}^{\mu\uparrow}(x) \Theta_{i_m,i_m+\tau'}^{\nu\downarrow}(x) \nonumber \\
&\times f_{i_n+\tau, i_m+\tau'},
\end{eqnarray}
where 
\{$\eta^{\mu\nu}_{{\tau} {\tau}'}$\} represent variational parameters.
 $\sum_{\tau (\tau')}$ is taken over ${\tau}$ (${\tau}'$) that satisfies the following condition : $0 \leq |\bm{\delta}|  \leq r^{\rm max}$, where $\bm{\delta} = \bm{r}_{i_n+\tau} - \bm{r}_{i_n}$ ($\bm{\delta} = \bm{r}_{i_m+\tau'}-\bm{r}_{i_m}$). 
We usually choose $r^{\rm max}$ as the range of hopping terms in the Hamiltonian. 
Here, 
we drop the electron indices $n$ and $m$ for simplicity of notation, 
and we define $\Theta_{i,i+\tau}^{\mu\sigma}(x)$ as 
\begin{eqnarray}
&&\Theta_{i,i+\tau}^{0\sigma}(x) = \delta_{i,i+\tau}, 
\\
&&\Theta_{i,i+\tau}^{1\sigma}(x) = \braket{D_{i} H_{i+\tau}}_x, 
\\
&&\Theta_{i,i+\tau}^{2\sigma}(x) =  \braket{n_{i\sigma} h_{i,-\sigma} n_{i+\tau,-\sigma} h_{i+\tau\sigma}}_x,
\\
&&\Theta_{i,i+\tau}^{3\sigma}(x) = \braket{D_{i} n_{i+\tau,-\sigma} h_{i+\tau\sigma}+n_{i\sigma} h_{i,-\sigma} H_{i+\tau}}_x, 
\end{eqnarray}
where $\delta_{i,i+\tau}$ represents the Kronecker delta, $h_{i\sigma} = 1-n_{i\sigma}$ and $\braket{A}_x = \braket{x | A | x}/\braket{x | x}$. 
We impose that $\eta^{\mu\nu}_{{\tau} {\tau}'}$ has the inversion symmetry and is independent of spin indices, namely $\eta^{\mu\nu}_{{\tau} {\tau}'} = \eta^{\mu\nu}_{-{\tau}, {\tau}'} = \eta^{\mu\nu}_{{\tau},-{\tau}'} = \eta^{\nu\mu}_{{\tau} {\tau}'}$. 
Furthermore, $\eta^{00}_{{\tau}{\tau}'}$ is replaced with 1 when $\Theta_{i,i+\tau}^{1\sigma}(x)=0$ for any $\tau$ and $\sigma$. 
Note that the introduction of backflow correlations make computational costs heavy because we need to recalculate the element of pairing wave function whenever we generate a candidate of the next sample and calculate expectation values of off-diagonal operators.

Finally, we explain the difficulties in operating the spin projection on the pair-product wave function with backflow correlations. From the definition of the pairing amplitudes with backflow correlations, the trial wave function with the spin projection and backflow correlations is described as
\begin{eqnarray}
&&\mathcal{L^S}\ket{\phi^b} 
=  \sum_x \ket{x}  \frac{2S+1}{2} \int d\beta \sin \beta P_S(\cos \beta) \braket{x | e^{i\beta S^y} |\phi^b} \nonumber \\
&&\propto \sum_{x,x'} \ket{x}  \int d\beta \sin \beta P_S(\cos \beta) \braket{x | e^{i\beta S^y} |x'} {\rm Pf} X^b(x'). 
\end{eqnarray}
Here, we need to take the summation over real space configurations $x'$ because the skey-symmetric matrix depends on $x'$. However, we cannot usually take this summation because the number of $x'$ grows exponentially as the system size increases.
Thus, it is difficult to operate the spin projection on a trial wave function with backflow correlations.
\section{Results}
In this section, we show the accuracy and efficiency of the t-VMC method. For benchmark tests, we consider the quench dynamics in fermionic Hubbard models. 
We compare the t-VMC results with numerically exact results obtained by calculating the formal solution of the time-dependent Schr\"odinger equation and time-dependent DMRG.

\subsection{Model and Setting}
For benchmark tests, we consider the fermionic Hubbard model which is defined as
\begin{eqnarray}
  \mathcal{H}(t) = -t_{\rm hop}\sum_{<i,j>,\sigma} c^\dagger_{i\sigma} c_{j\sigma} +U(t)\sum_i n_{i\uparrow}n_{i\downarrow},
\end{eqnarray}
where $t_{\rm hop}$ represents the hopping amplitude between nearest-neighbor sites and $U(t)$ represents the time-dependent onsite interaction.
In this paper, we set $t_{\rm hop}=1.0$.
In two-dimensional cases, we treat the square lattice.
We consider the linear-ramp quench where the strength of interaction $U(t)$ is linearly changed from $U_i$ to $U_f$ during time $t_q$:
\begin{eqnarray}
U(t) =  \left\{ \begin{array}{ll}
U_i + \frac{U_f-U_i}{t_q} t  & (0 \leq t \leq t_q) \\
U_f  & (t \geq t_q), \\
  \end{array} \right.
\end{eqnarray}
where $U_i$ and $U_f$ represent the strength of interaction before and after the quench protocol, respectively.

For equilibrium systems, there exist many theoretical studies on the Hubbard models\cite{JPSJ.75.114706,yokoyama2012crossover,PhysRevLett.108.216401,PhysRevB.86.241106,PhysRevB.90.115137}. 
In the present models, the ground state at half-filling is believed to be the antiferromagnetic insulator for any nonzero $U/t_{\rm hop}$ due to the Fermi surface nesting. 
Away from half-filling, this model shows rich phases such as correlated metals, antiferromagnetic metal and $d$-wave superconducting states\cite{yokoyama2012crossover,PhysRevLett.108.216401,PhysRevB.86.241106,PhysRevB.90.115137}. 
 According to the VMC calculations, such superconducting states appear above $U/t_{\rm hop} \gtrsim 6.0$ in an interval of the doping concentration\cite{PhysRevB.90.115137}. 

In all the t-VMC calculations, we impose a boundary condition so that the closed shell condition is satisfied for $U(t)=0.0$. We choose the discrete time step $\Delta t =1.0 \times 10^{-2}/U(t)$ for $U(t) > t_{\rm hop}$ and $\Delta t =1.0 \times 10^{-2}/t_{\rm hop}$ for $U(t) < t_{\rm hop}$ in the Runge-Kutta procedure.
\subsection{Comparison with Exact Results}

In this section, we mainly show the t-VMC results compared with ``exact" results.
Here, the ``exact" results mean those obtained by directly calculating the formal solution of the time-dependent Schr\"odinger equation:
\begin{eqnarray}
\ket{\psi(t+\Delta t)}
&=& e^{-i\mathcal{H}\Delta t} \ket{\psi(t)}\nonumber \\
&=& \sum_{n=0}^{M-1} \left( \frac{-i\mathcal{H}\Delta t}{n!} \right)^n \ket{\psi(t)} + \mathcal{O}(\Delta t^M),\quad 
\end{eqnarray}
where the wave function is completely expanded in the full Hilbert space. 
The errors in this calculation arise only from the time discretization with the amplitude of $\mathcal{O}(\Delta t^M)$. 
To preserve the total energy and the norm of the wave function with good accuracy, we have to choose a small time-grid $\Delta t$ and a high order $M$. 
In this study, we choose $\Delta t = 1.0 \times 10^{-3}$ and $M=3$ for systems at half filling and $\Delta t = 1.0 \times 10^{-2}$ and $M=5$ for doped systems.
Although the exact results can be obtained irrespective of the interaction strength, the system size is severely restricted to small systems because of the exponentially growing computational cost with the increase of the system size.
In order to check whether our trial wave function is accurate even for larger systems, we here compare the t-VMC results with time-dependent DMRG results which are expected to be highly accurate in the thermodynamic limit.

\begin{figure}[htbp]
  \begin{center}
   \includegraphics[width=83mm]{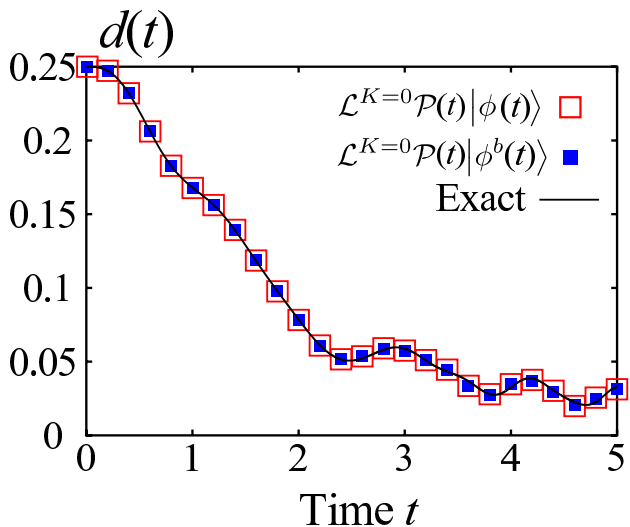}
  \end{center}
  \caption{(Color online) Time evolution of double occupancy $d(t)$ for linear-ramp quench in one-dimensional Hubbard model at half-filling. The parameter set is chosen as $(U_i,U_f,t_q,N_s)$=(0.0, 12.0, 5.0, 10). Red open squares represent the results computed by using the trial wave function with the momentum projection but without backflow correlations and blue solid squares represent the results obtained by using the trial wave function with backflow correlations and the momentum projection. Black curve represents the exact result. The range of backflow parameters $r^{\rm max}$ is chosen as $1$. To verify the effect of backflow correlations, we here optimized trial wave functions directly without the Monte Carlo integration.}
  \label{ramp5ns10_bf}
\end{figure}

In this section, we mainly show the results obtained by using a trial wave function with all quantum-number projections but without backflow correlations $\ket{\psi(t)} = \mathcal{L}^{\bm{K}=0} \mathcal{L}^{S=0} \mathcal{P}(t) \ket{\phi(t)}$.
Although we also tried a trial wave function with backflow correlations $\ket{\psi(t)} = \mathcal{L}^{\bm{K}=0} \mathcal{P}(t) \ket{\phi^b(t)}$ in our benchmarks,
we did not observe any clear improvement at least for small systems. 
Figure \ref{ramp5ns10_bf} shows an example of the time evolution of the double occupancy for the one-dimensional Hubbard model at half filling and $(U_i,U_f,t_q,N_s)$=(0.0, 12.0, 5.0, 10). 
This result indicates no effect of the backflow correlations. 
Note that backflow correlations may improve the description of dynamics in large systems with geometrical frustrations in a strong coupling region because these correlations are important for the description of ground states for such systems\cite{PhysRevB.78.041101,PhysRevB.83.195138}.

\subsubsection{One- and two-dimensional Hubbard model at half-filling}

In this section, we calculate the time evolution of several physical quantities. The physical quantities we have measured are the double occupancy $d(t)$, the momentum distribution $n(t;\bm{k})$ and the spin structure factor $S_s(t;\bm{q})$ defined as
\begin{eqnarray}
\left. \begin{array}{l}
  d(t)  =  {\displaystyle \frac{1}{N_s} \sum\limits_i^{N_s}} \braket{n_{i\uparrow} n_{i\downarrow}}, \\
   n(t; \bm{k}) =  {\displaystyle \frac{1}{2N_s}\sum_{i,j,\sigma}}\braket{c^\dagger_{i\sigma} c_{j\sigma}}e^{i\bm{k}\cdot (\bm{r}_i - \bm{r}_j)}, \\
  S_s(t;\bm{q})  = {\displaystyle \frac{1}{3N_s} \sum_{i,j}^{N_s}}\braket{\bm{S}_i \cdot \bm{S}_j}e^{i\bm{q}\cdot (\bm{r}_i - \bm{r}_j)},
  \end{array} \right.
\end{eqnarray}
respectively. Here, $\bm{k}$ and $\bm{q}$ are wavenumbers in the Brillouin zone. 
In addition, $\bm{S}_i=1/2 \sum_{\sigma,\sigma'} c^\dagger_{i\sigma} \bm{\sigma}_{\sigma,\sigma'}c_{i\sigma'}$, where $\bm{\sigma}$ are the Pauli matrices.

\begin{figure}[htbp]
  \begin{center}
   \includegraphics[width=83mm]{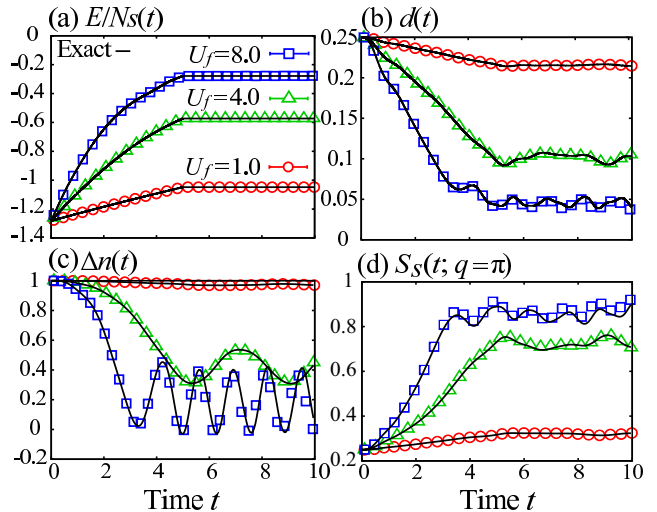}
  \end{center}
  \caption{(Color online) Time evolution of (a) energy per site $E/N_s(t)$, (b) double occupancy $d(t)$,  (c) jump of momentum distribution at Fermi energy $\Delta n(t) = n(t;q=\pi/2)-n(t;q=\pi/2+\pi/N_s)$, and (d) spin structure factor $S_s(t;q=\pi)$ for the linear-ramp quenches in one-dimensional Hubbard model at half filling. The parameter set is chosen as $(U_i,t_q,N_s)$=(0.0, 5.0, 16). Symbols and curves represent the t-VMC results and the exact results, respectively. Error bars indicate the statistical errors arising from the Monte Carlo sampling, but most of them are much smaller than the symbol sizes here and in the following figures.}
  \label{ramp5ns16}
\end{figure}

\begin{figure}[htbp]
  \begin{center}
   \includegraphics[width=83mm]{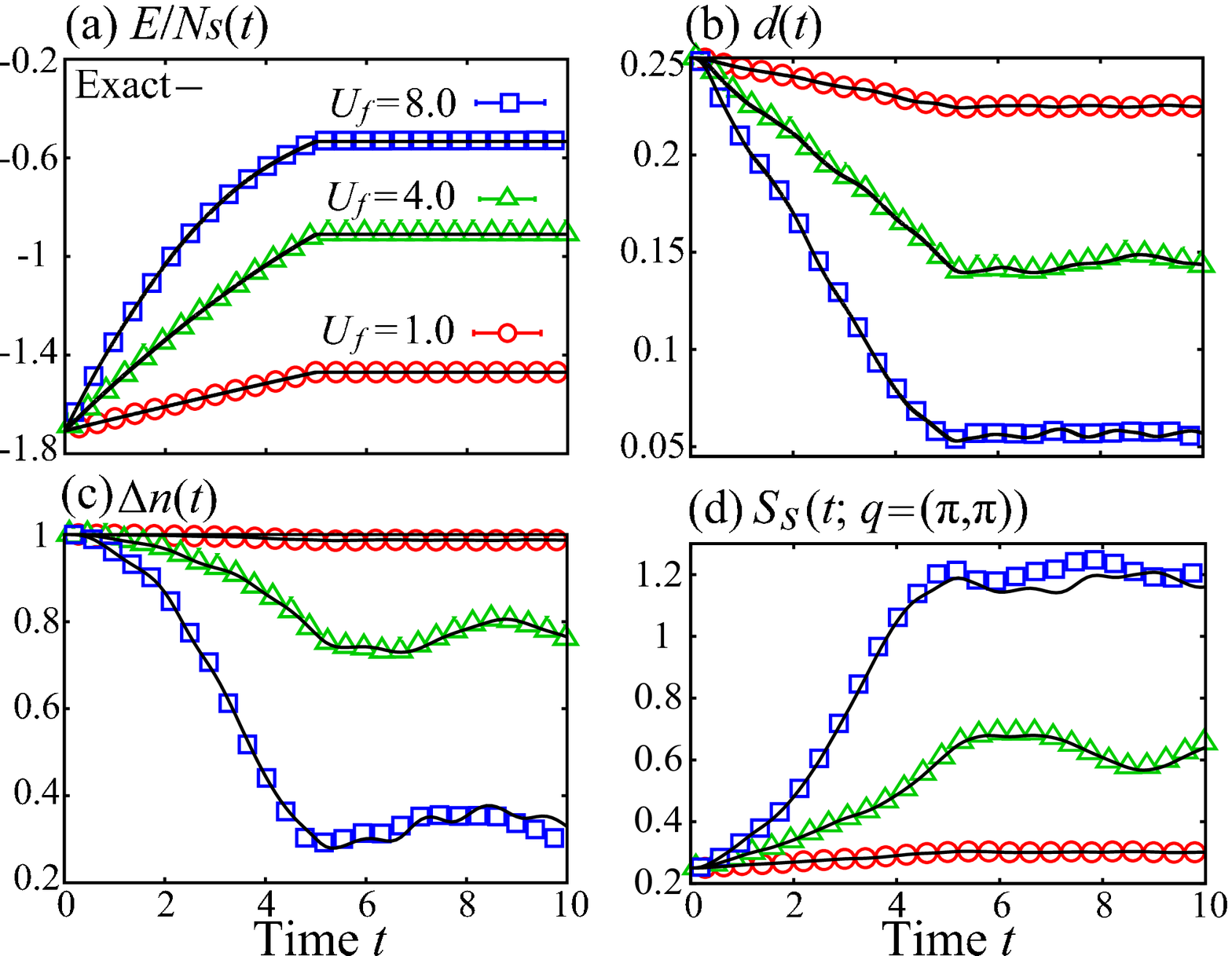}
  \end{center}
  \caption{(Color online) Time evolution of (a) energy per site $E/N_s(t)$, (b) double occupancy $d(t)$,  (c) jump of momentum distribution at Fermi energy $\Delta n(t) = n(t; (\pi,0)) - n(t; (\pi,\pi/N_s))$, and (d) spin structure factor $S_s(t;{\bm q}=(\pi,\pi))$ for the linear-ramp quenches in Hubbard model on square lattice at half filling. The parameter set is chosen as $(U_i,t_q,N_s)$=(0.0, 5.0, 4$\times$4). Symbols and curves represent the t-VMC results and the exact results, respectively.}
  \label{ramp5ns4x4}
\end{figure}

Figures \ref{ramp5ns16} and \ref{ramp5ns4x4} show the time evolution of several quantities in one dimension for $(U_i,t_q,N_s)$=(0.0, 5.0, 16) and in two dimensions for $(U_i,t_q,N_s)$=(0.0, 5.0, 4$\times$4), respectively.  The dimension of the Hilbert space in these systems is  $_{N_s}C_{N} \times _{N_s}C_{N}  \approx 10^8$.
However, our trial wave function has only several hundreds parameters.
Nevertheless, the t-VMC results well reproduce the exact results.
These results show that our trial wave function offers a highly accurate and efficient description of quantum dynamics for strongly correlated electron systems.

Here, we briefly comment on a dynamical transition in the Hubbard models at half-filling. 
Several works showed that the dynamics of the jump $\Delta n(t)$ is different depending on the strength of interaction\cite{Eckstein2009,Hamerla2013,PhysRevB.89.104301}.
For weak interactions, the jump $\Delta n(t)$ decreases gradually from $\Delta n(t)=1.0$ to a constant.
On the other hand, for strong interactions, the jump $\Delta n(t)$ exhibits a collapse-and-revival oscillation.
In Fig.\ref{ramp5ns16}(c), we observe clear collapse-and-revival oscillations in one-dimensional system especially at $U_f=8.0$. However, Fig.\ref{ramp5ns4x4}(c) shows that only weak oscillations are detected even at large $U_f$ in the two-dimensional system.
The difference appears to show a qualitative difference between one- and two-dimensional systems.
However, it is difficult to conclude whether a dynamical transition to the collapse-and-revival oscillation happens since the system sizes are too small to measure physical properties right at the Fermi surface, especially in two-dimensional case. 
To study nonequilibrium properties such as the dynamical transition, we need to treat larger system size. We leave its analysis for a future study.

\begin{figure}[htbp]
  \begin{center}
   \includegraphics[width=83mm]{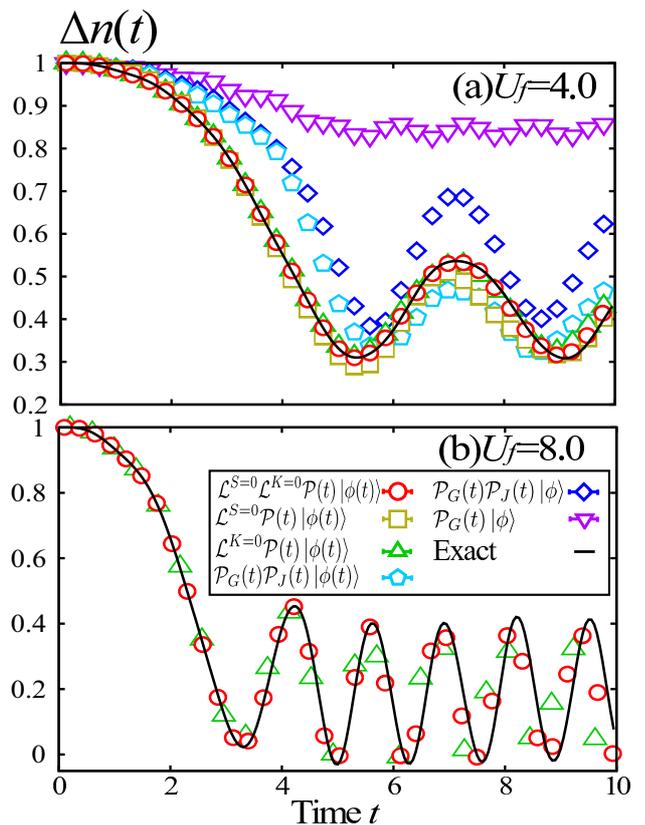}
  \end{center}
  \caption{(Color online) Time evolution of the jump of the momentum distribution near Fermi energy $\Delta n(t)$ for the linear-ramp quenches to (a)$U_f=4.0$ and (b)$U_f=8.0$ in one-dimensional Hubbard model at half-filling. The parameter set is chosen as $(U_i,t_q,N_s)$=(0.0, 5.0, 16). 
Symbols represent the t-VMC results obtained by using different trial wave functions. In the legend, time-dependent part of the trial wave functions include variational parameters we optimized. Solid curves represent the exact results.}
  \label{wf_ns16}
\end{figure}

Next we show the dependence on time-dependent trial wave functions. 
In Fig. \ref{wf_ns16}(a), we present the t-VMC results for several different time-dependent trial wave functions for the quench from $U_i=0.0$ to $U_f=4.0$. 
We note that $|\phi\rangle$ instead of $|\phi(t)\rangle$ represents the one-body part fixed through the time evolution at $|\phi(t=0)\rangle$ optimized in the ground state before quenching.
In two of the trial wave functions, we optimized all variational parameters with the momentum projection and the results agree with the exact ones. 
Other results are obtained without optimizing some part of variational parameters or without operating the momentum projection.
As seen in this figure, these results do not reproduce the exact results with substantial discrepancies. 
Especially, we find that the result obtained by optimizing only the Gutzwiller factor clearly disagrees with the exact one. 
The main reason for this disagreement is that Gutzwiller-type wave function cannot describe insulating states because of a lack of long-range off-site correlations which the Jastrow factor includes. 
In fact, the result obtained by optimizing only the Gutzwiller-Jastrow factor shows qualitative agreement with the exact one. This tendency is similar to that in equilibrium systems where a Gutzwiller-type wave function fails in reproducing the physical properties of the Hubbard models in finite dimensions\cite{yokoyama1987variational,PhysRevLett.94.026406}.
One might think that variational parameters in one-body part are not so important to describe the dynamics qualitatively because optimizing the one-body part affects only the amplitude of the oscillation after the quench in Fig.\ref{wf_ns16} (a). 
However, as we see clearly in Sec.\ref{sub_sc}, variational parameters in one-body part play an important role even in a qualitative description of nonequilibrium states.
In Fig. \ref{wf_ns16}(a), we do not see any improvements by operating the spin projection to $\mathcal{L}^{K=0}\mathcal{P}(t)\left| \phi(t) \right\rangle$.
However, for the quench to strong interaction $U_f=8.0$ in Fig. \ref{wf_ns16}(b), the difference between the two wave functions with the momentum projection is more obvious than that for $U_f=4.0$. 
These results suggest that it is better to operate both of the quantum-number projections on the trial wave function especially in the strong coupling region.
These results show that in order to obtain accurate results by the t-VMC method, we should operate the quantum-number projection and optimize all the variational parameters. 
This accuracy sensitive to trial wave functions is similar to that for ground states (see Ref.[\onlinecite{tahara2008variational,kaneko2013improved,PhysRevB.90.115137}] and Appendix.\ref{gs_vmc}).

\subsubsection{Size dependence}
In the previous subsection, we have shown the benchmark results for small systems at half-filling. 
In order to check the accuracy of our trial wave function for larger systems,
we here compare the t-VMC results with highly accurate DMRG result in the thermodynamic limit. 
Since the VMC method offers the results for finite-size systems, we need to investigate the system-size dependence of our results.

\begin{figure}[htbp]
  \begin{center}
   \includegraphics[width=85mm]{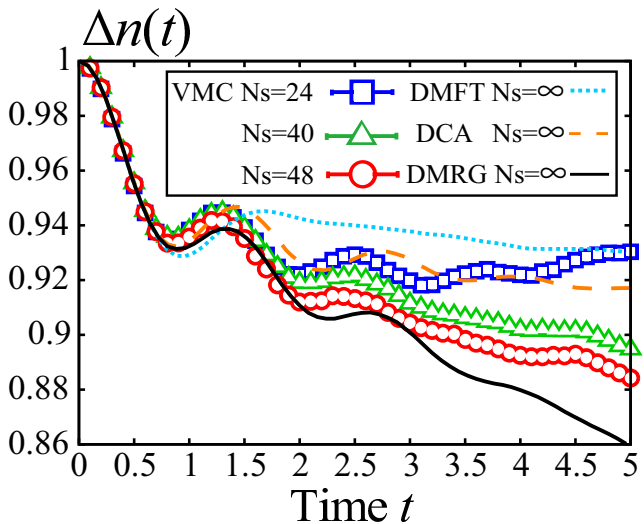}
  \end{center}
  \caption{(Color online) Time evolution of jump of momentum distribution near Fermi energy for sudden quench from $U_i=0.0$ to $U_f=1.0$ by using several methods. The DMRG, DMFT and DCA results are taken from Ref.[\onlinecite{Tsuji2014}]. 
Symbols, dotted line, dashed line and solid curve represent the results obtained by t-VMC, DMFT, DCA and DMRG, respectively. 
  The cluster-size employed in the DCA calculation is $N_c=64$.}
  \label{dnk_dmrg}
\end{figure}

In Fig. \ref{dnk_dmrg}, we present the t-VMC results of $\Delta n(t)$ for sudden quench protocol ($t_q=0.0$) from $U_i=0.0$ to $U_f=1.0$ in the one-dimensional Hubbard model at half-filling. 
In the t-VMC calculations, we operated only the momentum projection to reduce numerical cost since spin projection is not so important in the region of small interaction (See Fig. \ref{wf_ns16}).
To check the system-size dependence, we showed three t-VMC results with different system sizes.
For comparisons, we also show the results obtained by DMRG, DMFT and DCA\cite{Tsuji2014}. 
As an impurity solver of the DMFT and DCA calculations, the iterative perturbation theory (IPT) was employed. 
In the DCA calculation, the reciprocal wavevector $K$ satisfies the following condition: $K = 2n\pi/N_c$, where $n$ represents integer and $N_c$ represents a cluster size. Here, $N_c=64$ .
As seen in this figure, the jump $\Delta n(t)$ obtained by DMRG relaxes slowly with an oscillation. 
This feature is observed in both the DCA and t-VMC results although the DMFT result does not show the oscillation clearly after $t > 1.5$.
However, the DCA result shows a clear deviation from the DMRG result at long time even when large cluster size is used.
Tsuji and his coworkers have reported that this deviation comes from the quantum corrections from higher-order diagrams neglected in the IPT \cite{Tsuji2014}.
On the other hand, our t-VMC results have a strong size-dependence but approach the result of DMRG at long time as the system size increases. 
These results imply that our trial wave function in the t-VMC method have successfully included the quantum fluctuation beyond the DCA result.

\subsubsection{Superconducting correlation in hole-doped Hubbard model in two dimensions}
\label{sub_sc}
The doped Hubbard model on the square lattice is one of the simplest models for studying the high-$T_c$ superconductivity in copper oxides. In this model for equilibrium, many theoretical studies have proposed the existence of $d$-wave superconducting states in the hole-doped region with strong on-site interaction\cite{yokoyama2012crossover,PhysRevLett.108.216401,PhysRevB.86.241106,PhysRevB.90.115137}.
Recently, some experiments have reported that nonequilibrium superconducting states in copper oxides have been realized even at room temperature\cite{Hu2014, Kaiser2014}. 
In order to theoretically determine whether time-evolved states realize superconducting state, it is necessary to calculate time evolutions of pairing correlation functions.

As a benchmark for the $d_{x^2-y^2}$-wave pairing correlation functions $P_{d}(t;\bm{r})$, we consider the hole-doped Hubbard models in two dimensions. The pairing correlation function is defined as
\begin{eqnarray}
  P_{d}(t;\bm{r}) &=& \frac{1}{2N_s}\sum_{\bm{r}_i}^{N_s} \left[ \braket{\Delta_d^\dagger(\bm{r}_i) \Delta_d(\bm{r}_i+\bm{r})}\right. \nonumber  \\ 
  && \qquad \qquad \quad \quad \left.+ \braket{\Delta_d(\bm{r}_i) \Delta_d^\dagger(\bm{r}_i+\bm{r})}\right]. \qquad
\end{eqnarray}
Here, $\Delta_d(\bm{r}_i)$ represents the $d_{x^2-y^2}$-wave superconducting order parameter defined as
\begin{eqnarray}
  \Delta_d(\bm{r}_i) = \frac{1}{\sqrt{2}} \sum_{j} f_{d}(\bm{r}_j-\bm{r}_i) (c_{i\uparrow}c_{j\downarrow} - c_{i\downarrow}c_{j\uparrow}), 
\end{eqnarray}
where 
\begin{eqnarray}
  f_{d}(\bm{r}) = \delta_{r_y,0}(\delta_{r_x,1}+\delta_{r_x,-1})-\delta_{r_x,0}(\delta_{r_y,1}+\delta_{r_y,-1})\ \ \ 
\end{eqnarray}
is the form factor which describes the $d_{x^2-y^2}$-wave symmetry and $\bm{r} = (r_x, r_y)$.

\begin{figure}[htbp]
  \begin{center}
   \includegraphics[width=83mm]{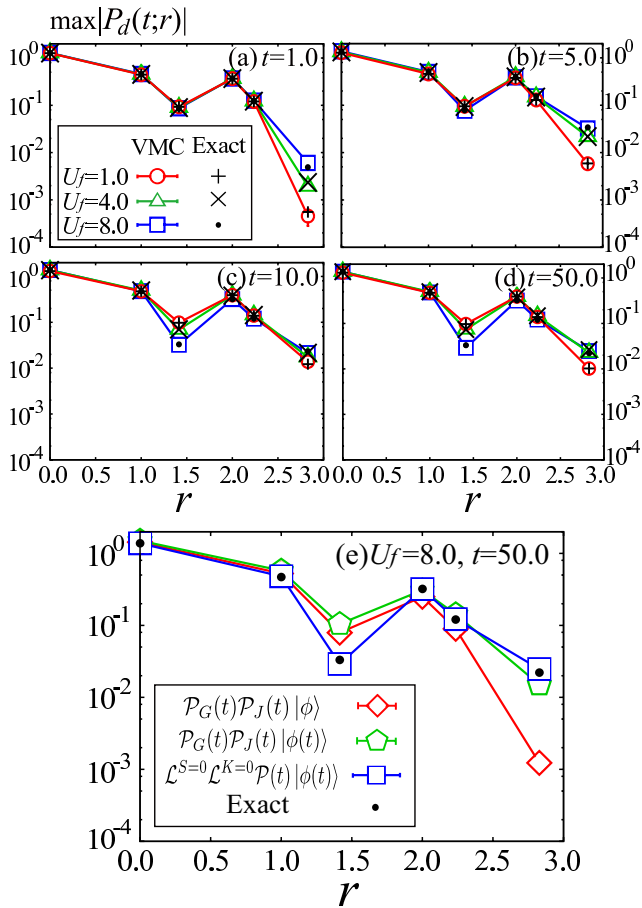}
  \end{center}
  \caption{(Color online) Time evolution of superconducting correlation functions ${\rm max}|P_{d}(t;r)|$ for the linear-ramp quenches in two-dimensional doped Hubbard model on square lattice. The parameter set is chosen as $(U_i,t_q,N_s, N)$=(0.0, 10.0, 4$\times$4, 12). The results are measured at time (a) $t=1.0$ (after quench protocol starts) , (b) $t=5.0$ (in the middle of quench protocol), (c) $t=10.0$  (after quench protocol ends) and (d) $t=50.0$ (in long-time limit). 
  The dependence on trial wave functions is also shown in (e) for $U_f=8.0$ at $t=50.0$. The lattice spacing is used as a unit of distance. Open symbols represent the t-VMC results and the other symbols represent the exact results.
  The accurate results are obtained only when we use $\ket{\psi(t)} = \mathcal{L}^{\bm{K}=0} \mathcal{L}^{S=0} \mathcal{P}(t) \ket{\phi(t)}$.}
  \label{sc_ramp10ns4x4}
\end{figure}

In Figs. \ref{sc_ramp10ns4x4}(a)-(d), we compare the t-VMC results with the exact results of ${\rm max}|P_{d}(t;r)|$ at four different $t$'s in the linear-ramp quench from $U_i=0.0$ with $t_q = 10.0$. 
Here, ${\rm max}|P_{d}(t;r)|$ denotes the maximum absolute value of pairing correlation functions $|P_{d}(t;\bm{r})|$ among the same $r=|\bm{r}|$. 
As shown in Figs. \ref{sc_ramp10ns4x4}(a)-(d), our t-VMC results show good agreements with the exact results for all the distances at each time.
This accuracy of the superconducting correlations is the same as that for the ground state (see Appendix.\ref{gs_vmc}).

Figure \ref{sc_ramp10ns4x4}(e) shows the dependence of ${\rm max}|P_{d}(t;r)|$ on trial wave functions at long time $t=50.0$ for $U_f=8.0$.
We again note that $|\phi\rangle$ instead of $|\phi(t)\rangle$ represents the one-body part fixed through the time evolution at $|\phi(t=0)\rangle$ optimized in the ground state before quenching.
As seen in Fig. \ref{sc_ramp10ns4x4}(e), only the correlation function at the largest distance obtained by using $\mathcal{P}_G(t) \mathcal{P}_J(t) \left| \phi \right\rangle$ shows a large deviation from the exact result, i.e., its value is one order of magnitude lower than the other ones.
 This result implies that only optimizing the correlation factors is insufficient in reproducing pairing correlations in time evolution.
 In fact, by optimizing the amplitudes of singlet pairings $f_{ij}$ in one-body part in time evolution, 
 the t-VMC result at the largest distance well reproduces the exact one. 
It is important to obtain the long-range part of the ${\rm max}|P_{d}(t;r)|$ accurately because it enables us to detect the emergence of the superconducting phase in large systems. 
Therefore, to describe different nonequilibrium states flexibly, it is crucial to optimize the one-body part.

From these benchmark results, even the superconducting correlation can be well reproduced by using our trial wave function, which may inspire studies along this line in the t-VMC method. 
One of the intriguing studies is on the influence of strong laser pulse on superconductivity in correlated electron systems. 
Some recent works have shown that the hopping amplitude is reduced by applying strong laser and the relative strength of interaction to the transfer become effectively larger than before\cite{Tsuji2011,JPSJ.83.024706,ishikawa2014optical}. 
By using this effect, the $d$-wave superconductivity may grow because, for ground states, it grows as the on-site interaction increases.
However, in order to confirm whether nonequilibrium states show a true long-range order or not, calculations for larger systems are required.
Studies on nonequilibrium superconducting states in larger systems will be reported elsewhere.

\section{Summary and Outlook}
In summary, we have developed a time-dependent variational Monte Carlo method (t-VMC) for strongly correlated electron systems and have shown the benchmark results for the fermionic Hubbard model out of equilibrium. 
By comparing our t-VMC results with the exact results, we found that our trial wave function well reproduces exact time evolutions in both one and two dimensions. These results show that our trial wave function offers an accurate and efficient description of nonequilibrium states in strongly correlated electron systems.

One of the advantages of the VMC method is its wide applicability. In fact, the VMC method can be applied to complicated {\it ab initio} effective models derived by the downfolding scheme and contributed to identifying mechanism of physical properties in real materials\cite{shinaoka2012mott,misawa2014superconductivity}. 
Applications of the t-VMC method to such models are intriguing future issues. Furthermore, the VMC method can be applied not only to purely electronic systems but also to electron-phonon coupled systems\cite{ohgoe2014variational}. Therefore, it would be possible to study the phenomena of relaxation process and photoinduced phase transitions through phonon modes in real materials such as copper oxides.

\begin{acknowledgments}
The code was developed based on the VMC code implemented for electron systems with contributions by D. Tahara and S. Morita. The authors thank T. Misawa  for useful comments on the present study and for providing them with the code for solving time-dependent Schr\"odinger equation. In the code for solving time-dependent Schr\"odinger equation, some routines in TITPACK version 2 was partly used. K.I. also thanks Y. Yamaji and K. Takai for fruitful discussions. 
The authors thank the Supercomputer Center, the Institute for Solid State Physics, the University of Tokyo for the facilities. 
This work was financially supported by Japan Society for the Promotion of Science through Program for Leading Graduate Schools (MERIT), 
the MEXT HPCI Strategic Programs for Innovative Research (SPIRE), RIKEN Advanced Institute for Computational Science (AICS) through the HPCI System Research Project (under Grants No. hp130007, hp140215 and hp150211) and the Computational Materials Science Initiative (CMSI).
This work was also supported by a Grant-in-Aid for Scientific Research (No. 22104010 and No. 22340090) from Ministry of Education, Culture, Sports, Science and Technology, Japan.  
\end{acknowledgments}
%
%
\appendix
\section{Time-Dependent Variational Principle for Norm-Independent Dynamics}
\label{detail}

In this appendix, we review the TDVP for norm-independent dynamics in terms of the principle of least action\cite{PhysRevB.88.075133,kramer1981geometry}. 
We apply the principle of least action to an action $S=\int dt L(\bm{\overline{\alpha}}, \bm{\alpha})$ on the manifold $\mathcal{M}$ of a trial wave function $\ket{\psi_{\bm{\alpha}}}$. Here, the Lagrangian $L(\bm{\overline{\alpha}}, \bm{\alpha})$ is described as
\begin{eqnarray}
L(\bm{\overline{\alpha}}, \bm{\alpha}) &=& \frac{i}{2} \left( \braket{\dot{\psi}_{\bm{\overline{\alpha}}} | \psi_{\bm{\alpha}}} - \braket{\psi_{\bm{\overline{\alpha}}} | \dot{\psi_{\bm{\alpha}}}} \right) - \braket{\psi_{\bm{\overline{\alpha}}} | \mathcal{H} | \psi_{\bm{\alpha}}}, \nonumber
\end{eqnarray}
where $\bm{\alpha}$ and $\overline{\bm{\alpha}}$ represents variational parameters and its complex conjugates, respectively.
Although the norm of the wave function is preserved under the exact time evolution, the evolution on the  manifold $\mathcal{M}$ may break the norm-conservation. 
To remove the restriction on the norm, the norm-dependent Lagrangian $L(\bm{\overline{\alpha}}, \bm{\alpha})$ should be normalized. Thus, the modified Lagrangian $\tilde{L}(\bm{\overline{\alpha}}, \bm{\alpha})$ for norm-independent dynamics is introduced: 

\begin{eqnarray}
\tilde{L}(\bm{\overline{\alpha}}, \bm{\alpha}) &=& L(\bm{\overline{\alpha}}, \bm{\alpha})/\braket{\psi_{\bm{\overline{\alpha}}} | \psi_{\bm{\alpha}}} \nonumber \\
&=& \frac{i}{2} \frac{\braket{\dot{\psi}_{\bm{\overline{\alpha}}} | \psi_{\bm{\alpha}}} - \braket{\psi_{\bm{\overline{\alpha}}} | \dot{\psi_{\bm{\alpha}}}}}{\braket{\psi_{\bm{\overline{\alpha}}} | \psi_{\bm{\alpha}}}} - \frac{\braket{\psi_{\bm{\overline{\alpha}}} | \mathcal{H} | \psi_{\bm{\alpha}}}}{\braket{\psi_{\bm{\overline{\alpha}}} | \psi_{\bm{\alpha}}}}. \nonumber
\end{eqnarray}
The variation of the corresponding action $\delta \tilde{S}$ with respect to variations of parameters $\bra{\psi_{\bm{\overline{\alpha}}}} \rightarrow \bra{\psi_{\bm{\overline{\alpha}}}} + \bra{\delta \psi_{\bm{\overline{\alpha}}}}$ is given by
\begin{eqnarray}
\delta \tilde{S}(\bm{\overline{\alpha}}, \bm{\alpha}) 
&=&  \int dt \frac{\braket{ \delta \psi_{\bm{\overline{\alpha}}} | i\left( \frac{d}{dt} -\frac{\braket{\psi_{\bm{\overline{\alpha}}} | \dot{\psi_{\bm{\alpha}}}}}{\braket{\psi_{\bm{\overline{\alpha}}} | \psi_{\bm{\alpha}}}} \right) | \psi_{\bm{\alpha}}}}{\braket{\psi_{\bm{\overline{\alpha}}} | \psi_{\bm{\alpha}}}} \nonumber \\
&&-\int dt \frac{\braket{ \delta \psi_{\bm{\overline{\alpha}}} | \left( \mathcal{H} - \frac{\braket{\psi_{\bm{\overline{\alpha}}}| \mathcal{H} |\psi_{\bm{\alpha}}}}{\braket{\psi_{\bm{\overline{\alpha}}} | \psi_{\bm{\alpha}}}} \right) | \psi_{\bm{{\alpha}}}}}{\braket{\psi_{\bm{\overline{\alpha}}} | \psi_{\bm{\alpha}}}} \nonumber \\
&=& \int dt \frac{\braket{ \delta \psi_{\bm{\overline{\alpha}}} | \left(1-\frac{\ket{\psi_{\bm{\alpha}}} \bra{\psi_{\bm{\overline{\alpha}}}}}{\braket{\psi_{\bm{\overline{\alpha}}}| \psi_{\bm{\alpha}}}} \right) \left[ i\frac{d}{dt} - \mathcal{H} \right] | \psi_{\bm{\alpha}}}}{\braket{\psi_{\bm{\overline{\alpha}}}| \psi_{\bm{\alpha}}}}. \nonumber
\end{eqnarray}
Stationarity of the modified action $\delta \tilde{S} = 0$ leads to the variational equation on the manifold $\mathcal{M}$
\begin{eqnarray}
\braket{ \delta \psi_{\bm{\overline{\alpha}}} | \left(1-\frac{\ket{\psi_{\bm{\alpha}}} \bra{\psi_{\bm{\overline{\alpha}}}}}{\braket{\psi_{\bm{\overline{\alpha}}}| \psi_{\bm{\alpha}}}} \right) \left[ i\frac{d}{dt} - \mathcal{H} \right] | \psi_{\bm{\alpha}}} = 0. \label{var}
\end{eqnarray}
Thus, the condition of minimizing the modified action for norm-independent dynamics is equivalent to Eq. (\ref{tdvp}) in the full Hilbert space. Based on Eq.(\ref{tdvp}) or Eq.(\ref{var}), we can easily derive the Euler-Lagrange equation described as
\begin{eqnarray}
\dot{\alpha_k} =\frac{d \alpha_k}{dt}= -i \sum_l^{N_p} (S^{-1})_{kl}g_l, \label{general}
\end{eqnarray}
where a matrix $S$ and a vector $g$ are described as
\begin{eqnarray}
&S_{kl} = \frac{\partial}{\partial \overline{\alpha_k}} \frac{\partial}{\partial \alpha_l}\ln \braket{\psi_{\bm{\overline{\alpha}}} | \psi_{\bm{\alpha}}} , \
& g_k = \frac{\partial}{\partial \overline{\alpha_k}} \frac{\braket{\psi_{\bm{\overline{\alpha}}} | \mathcal{H} | \psi_{\bm{\alpha}}}}{\braket{\psi_{\bm{\overline{\alpha}}} | \psi_{\bm{\alpha}}}}. \nonumber
\end{eqnarray}
\section{Dependence on Trial Wave Functions for Ground States of Hubbard Models}
\label{gs_vmc}

In this appendix, to gain insight into the dependence of physical properties on trial wave functions in the nonequilibrium states, we compare them with the benchmarks for the ground states of the Hubbard models.

\begin{figure}[htbp]
  \begin{center}
   \includegraphics[width=83mm]{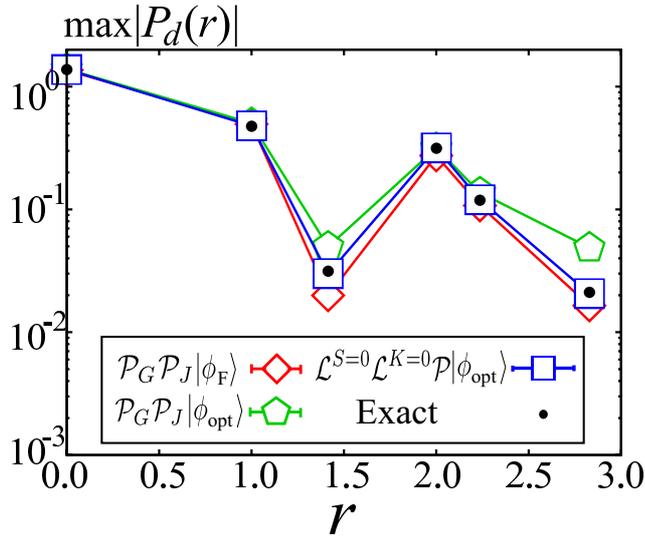}
  \end{center}
  \caption{(Color online) Superconducting correlation functions ${\rm max}|P_{d}(r)|$ in the ground state of two-dimensional doped Hubbard model on square lattice for $U/t_{\rm hop}=8.0, n=12/16$. 
  Here, $\mathcal{P}=\mathcal{P}_G \mathcal{P}_J$.
  The lattice spacing is used as a unit of distance. Open symbols represent the VMC results and the dots represent the ED results.}
  \label{sc_u8n6}
\end{figure}

Tables \ref{table:hf} and \ref{table:doped} show how physical properties depend on trial wave functions in the ground state of the Hubbard models at and away from half-filling, respectively. 
To show the accuracy of our VMC results, we also show the results obtained by the exact diagonalization (ED) method.
In these Tables, $\left| \phi_{\rm F} \right\rangle$ and $\left| \phi_{\rm opt} \right\rangle$ represent the Fermi sea state and optimized pair-product wave function, respectively. 
In all the cases in Table \ref{table:hf}, there are large discrepancies, especially in the jump $\Delta n$, between the results obtained by using the Gutzwiller-type wave function (GWF) $\mathcal{P}_G \left| \phi_{\rm F} \right\rangle$ and those of the ED method. 
The main reason for this is that GWF cannot describe insulating states as we described in Sec.\ref{twf}. 
By operating the Jastrow factor, the VMC results at half-filling are qualitatively consistent with those of the ED method.
For both half-filled and hole doped models, the energies obtained by using our best trial wave function $\mathcal{L}^{K=0}\mathcal{L}^{S=0} \mathcal{P}_G\mathcal{P}_J\left| \phi_{\rm opt}\right\rangle$ agree with those of the ED method, and the relative errors are less than 0.5\%.
Figure \ref{sc_u8n6} shows the pairing correlations obtained by using different trial wave functions in the doped Hubbard model on the square lattice for $U/t_{\rm hop} = 8.0, n=12/16$. 
As shown in Fig.\ref{sc_u8n6}, $\mathcal{L}^{K=0}\mathcal{L}^{S=0} \mathcal{P}_G\mathcal{P}_J \left| \phi_{\rm opt}\right\rangle$ has the best accuracy of the $d_{x^2-y^2}$-wave superconducting correlations for all the distances.
These trends on trial wave functions are similar to those in the t-VMC method.
\begin{table}[hbtp]
  \caption{
    Comparison of physical quantities obtained by the exact diagonalization (ED) method with those by using different trial wave function for one-dimensional Hubbard model at half-filling. 
  $E/N_s$, $\Delta n$ and $S_s(\pi)$ represents energy per site, jump of the momentum distribution near the fermi energy, and the spin structure factor, respectively. 
  The numbers in parentheses denote the statistical errors in the last digits.}
  \label{table:hf}
  \centering
  \begin{tabular}{lccc}
   \hline
       & $E/N_s$ &  $\Delta n$ & $S_s(\pi)$ \\
   \hline \hline
    $U/t_{\rm hop}$=4.0, $N_s$=16 &&&\\
    $\mathcal{P}_G \left| \phi_{\rm F}\right\rangle$ & -0.5280(5)  & 0.843(1) & 0.490(2) \\
    $\mathcal{P}_G\mathcal{P}_J \left| \phi_{\rm F}\right\rangle$ & -0.555(4)  & 0.5257(17) & 0.677(4)\\
    $\mathcal{P}_G\mathcal{P}_J \left| \phi_{\rm opt} \right\rangle$ & -0.5674(4)  & 0.3868(17) & 0.6878(23) \\
    $\mathcal{L}^{S=0}\mathcal{L}^{K=0}\mathcal{P}_G\mathcal{P}_J \left| \phi_{\rm opt} \right\rangle$ & -0.57605(1)  & 0.4220(5) & 0.7329(4)\\
    Exact(ED) & -0.57660  & 0.4326 & 0.7277\\
    \\
    $U/t_{\rm hop}$=8.0, $N_s$=16 &&&\\
    $\mathcal{P}_G \left| \phi_{\rm F}\right\rangle$ & -0.217(2)  & 0.447(3) & 0.813(5) \\
    $\mathcal{P}_G\mathcal{P}_J \left| \phi_{\rm F}\right\rangle$ & -0.3170(4)  & 0.168(3) & 0.904(3)\\
    $\mathcal{P}_G\mathcal{P}_J \left| \phi_{\rm opt} \right\rangle$ & -0.3238(3)  & 0.150(1) & 0.898(3) \\
    $\mathcal{L}^{S=0}\mathcal{L}^{K=0}\mathcal{P}_G\mathcal{P}_J \left| \phi_{\rm opt} \right\rangle$ & -0.32857(2)  & 0.1546(3) & 0.9561(7)\\
    Exact(ED) & -0.32904  & 0.1578 & 0.9556\\
    \hline
  \end{tabular}
\end{table}
\begin{table}[hbtp]
  \caption{
    Comparison of energy per site obtained by the ED method with those by using different trial wave function for doped Hubbard model on square lattice for $U/t_{\rm hop}$=8.0 at $n$=12/16. 
  The numbers in parentheses denote the statistical errors in the last digit.}
  \label{table:doped}
  \centering
  \begin{tabular}{lc}
   \hline
       & $E/N_s$  \\
   \hline \hline
    $\mathcal{P}_G\mathcal{P}_J \left| \phi_{\rm F}\right\rangle$ & -0.9373(1)  \\
    $\mathcal{P}_G\mathcal{P}_J \left| \phi_{\rm opt} \right\rangle$ &  -0.9452(2) \\
    $\mathcal{L}^{S=0}\mathcal{L}^{K=0}\mathcal{P}_G\mathcal{P}_J \left| \phi_{\rm opt} \right\rangle$ & -0.9728(1) \\
    Exact(ED) & -0.9774  \\
    \hline
  \end{tabular}
\end{table}


\bibliographystyle{prsty}
\bibliography{reference}
\end{document}